\documentclass[reprint,amsmath,amssymb,aps]{revtex4-1}

\usepackage{graphicx}
\usepackage{dcolumn}
\usepackage{bm}
\usepackage[T2A]{fontenc}
\usepackage[utf8]{inputenc}
\usepackage{color}

\newcommand\ket[1]{|#1\rangle}

\begin{document}

\title{Sub-shot-noise-limited fiber-optic quantum receiver}

\author{M.L.~Shcherbatenko}
\affiliation{Moscow State Pedagogical University, 1/1 M.Pirogovskaya St., Moscow 119992, Russia}

\author{M.S.~Elezov}
\affiliation{Moscow State Pedagogical University, 1/1 M.Pirogovskaya St., Moscow 119992, Russia}

\author{D.V.~Sych}\thanks{denis.sych@gmail.com}
\affiliation{P.N. Lebedev Physical Institute, Russian Academy of Sciences, 53 Leninskiy Prospekt, Moscow 119991, Russia}
\affiliation{QRate LLC, Novaya av. 100, Moscow 121353, Russia}
\affiliation{Moscow State Pedagogical University, 1/1 M.Pirogovskaya St., Moscow 119992, Russia}

\author{G.N.~Goltsman}\thanks{goltsman@rplab.ru}
\affiliation{Moscow State Pedagogical University, 1/1 M.Pirogovskaya St., Moscow 119992, Russia}

             
\begin{abstract}
We experimentally demonstrate a quantum receiver based on Kennedy scheme for discrimination between two phase-modulated weak coherent states. The receiver is assembled entirely from the standard fiber-optic elements and operates at the conventional telecom wavelength 1.55 microns. The local oscillator and the signal are transmitted through different optical fibers, and the displaced signal is measured with a high-efficiency superconducting nanowire single-photon detector. We show the discrimination error rate two times below the shot-noise limit.


\end{abstract}

\maketitle

\section*{\label{sec:level1} Introduction}

Since the beginning of telecommunications there has been an ever increasing demand to transmit more data. Optical signals are ideal carriers of information over long distances due to their high capacity and speed. An optical receiver has to convert the optical signal into the logical data. An important performance index of different receivers is the error rate, i.e the probability to transmit a single bit incorrectly. Quantum properties of light set the fundamental limits on the minimum error rate. With the use of Gaussian operations, the minimum error is bounded by the shot noise or so-called the standard quantum limit (SQL) \citep{Bergou:10}. The use of non-Gaussian operations (e.g. single-photon detectors) can reduce the error rate to a lower value, bounded from below by the Helstrom bound \citep{Helstrom:67, Bakut:68}. In theory, several types of quantum receivers may be used to overcome the SQL \citep{Kennedy:73, Becerra:11, Becerra:13, Bondurant:93, Muller:15, Tsujino:11, Tsujino:10, Chen:12, Wittmann:08, Takeoka:08, Guha:11, DiMario:19} or even approach the Helstrom bound for certain types of signals \citep{Dolinar:73, Sych:16, Takeoka:05, Sasaki:96}. However in practice, it is very difficult to experimentally demonstrate the sub-SQL performance due to several reasons.

First, most theoretical proposals of sub-SQL receivers involve optical displacement of the communication signal (i.e. interference of the signal with a reference light beam, called local oscillator (LO)) followed by its measurement with help of a single photon detector (SPD). To realize stable and high-contrast interference, LO must be perfectly mode-matched to the signal. Moreover, in the first theoretical proposal of the quantum receiver that reaches the Helstrom bound (Dolinar receiver)\citep{Dolinar:73}, the phase and amplitude of LO must be dynamically adjusted depending on the output of the SPD via an instantaneous feed-back. A recently developed optimal multi-channel quantum receiver \citep{Sych:16} does not require instantaneous feed-back, but involves several optical displacement operations. Another approach to reach the Helstrom bound involves non-linear transformation of the signal, which is extremely challenging to realize in practice \citep{Sasaki:96}. 

Second, the overall system detection efficiency of the receiver, including the quantum efficiency of SPD, must be high --- above 60\% for the binary signal \citep{Helstrom:67}. Apart from high quantum efficiency, the SPD must have low dark count rate and short dead time, which is rather difficult to combine. Superconducting nanowire single-photon detector (SNSPD)\citep{Goltsman:01} show great potential in this context, since they have quantum efficiency close to 100\%, dead time is below 10 ns, and dark counts can be as low as 0.01 counts per second \cite{Smirnov:18, Smirnov:15, Marsili:13, Natarajan:12}. SNSPD is coupled with single-mode fiber, which makes it relatively easy to integrate into a fiber-optic receiver scheme.

In this work we experimentally realize the simplest and historically the first sub-SQL receiver design, called Kennedy receiver \citep{Kennedy:73}. In order to meet the above-mentioned requirements for sub-SQL performance, we combine the standard single-mode fiber optics components to realize the high-extinction optical displacement operation and an SNSPD to detect the displaced signal. The receiver operates at the conventional telecommunication wavelength 1.55 microns. We experimentally demonstrate that the maximal improvement of the error rate is about 2 times with respect to the SQL.

Note that all previous experimental attempts towards the sub-SQL quantum receivers, as well as quantum receivers aimed at other optimization strategies \citep{Wittmann:10prl, Wittmann:10pra, Izumi:16}, were partially or fully implemented using optical components in free space. The use of standard fiber-optic components brings this area of research closer to real-world applications, since the vast majority of devices in classical optical communication use the optical fiber technology. Compared to free-space optics, fiber-optic devices are more compact, reproducible and convenient for practical use. It is easier to scale them up, which is important for the development of improved quantum receivers \cite{Takeoka:05,Sych:16}. 

\color{black}
\section{\label{sec:level2} Kennedy receiver}
In the simplest form, the problem of minimum--error state discrimination is as follows. Consider the signal  prepared in one of the two equiprobable phase modulated coherent states $\ket{\alpha}$,$\ket {-\alpha}$ --- the so-called binary phase-shift keyed (BPSK) signal. The main task of a receiver is to make some measurement of the signal and find out the actual signal state.

\begin{figure}[h]
\includegraphics[width=83mm] {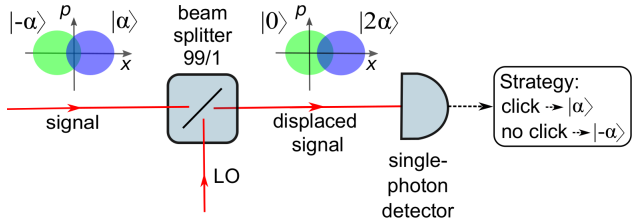}
\centering\caption{\label{fig:Kennedy} 
Schematics of a displacement-based Kennedy receiver. The mode-matched signal and local oscillator are combined on a 99:1 beam splitter such that after the beam splitter one of the possible states (e.g. $\ket{-\alpha}$) vanishes out due to destructive interference with the local oscillator (LO). The result of interference between the signal and LO is registered by a single-photon detector.}
\end{figure}

According to Kennedy's proposal \cite{Kennedy:73}, the measurement should be done by interfering the signal with the mode-matched reference beam (local oscillator, LO) on a beam splitter, such that after the beam splitter one of the possible states vanishes out due to destructive interference with the LO: ${\ket{\alpha},\ket{-\alpha}}\rightarrow {\ket{2\alpha},\ket{0}}$. To reduce attenuation of the signal, interference is performed on an almost completely transparent beam splitter (99:1) \cite{wallentowitz:96}. Such operation is called the exact nulling displacement. Afterwards, the displaced signal is measured with help of a single-photon detector (see Fig.~\ref{fig:Kennedy}). If the detector gives a photocount (``click'' event), then we can be sure that the signal was in the state which is not nulled (receive logical ``1''), while in the case of no photocounts (``no--click'' event), the most probable state is the nulled signal (receive logical ``0''). 

\begin{figure}[h]
\includegraphics[width=87mm]{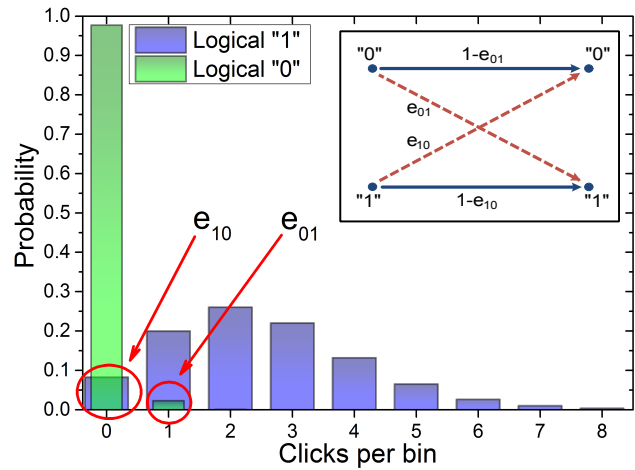}
\centering\caption{\label{fig:Error} 
Probability distribution of photo-counts for logical “0” (green histogram), and logical “1” (blue histogram). Discrimination errors correspond to the first component of the blue histogram (zero number of photo-counts) and all-but-first components of the green histogram (non-zero number of photons). The inset shows the error diagram.}
\end{figure}

The error diagram of the Kennedy receiver shown in Fig.~\ref{fig:Error} (inset). Errors $e_{10}$ of the ideal Kennedy receiver come from the fact that ``no--click'' events may appear not only from the nulled signal $\ket{0}$, but also from the zero-photon component of the signal $\ket{2\alpha}$, i.e. non-zero signal is decoded incorrectly upon a  ``no--click'' event. The probability distribution of $n-$photon components of the coherent state $\ket{x}$ is given by Poisson distribution $P_n(x)=e^{-x^2} x^{-2n}/n!$, which is shown in Fig.~\ref{fig:Error}.
Thus the error rate of the Kennedy receiver is given by 
\begin{equation}\label{eq:Ken} 
e_K=0.5 P_0(2\alpha)=0.5 e^{-4\,m},
\end{equation}
where $m=\alpha^2$ is the average number of photons in the initial signal.
 For comparison, the error rate of the heterodyne receiver is given by \cite{Sych:16}
\begin{equation}\label{eq:SQL} 
e_{SQL}=\frac{1}{2} \* (1-Erf(\sqrt{2 \* m})),
\end{equation}
where $erf(x)=\frac{2}{\sqrt{\pi}} \* \int_0^x e^{-t^2}dx$, and the Helstrom bound is given by 
 \begin{equation}\label{eq:Hel} 
e_{Hel}=\frac{1}{2}\*(1-(\sqrt{1-e^{-4\,m}})).
\end{equation}
These values, normalized to SQL, are shown by solid and dotted lines in Fig.~\ref{fig:Graphics}. 

\begin{figure*}[ht]
\includegraphics[width=160mm]{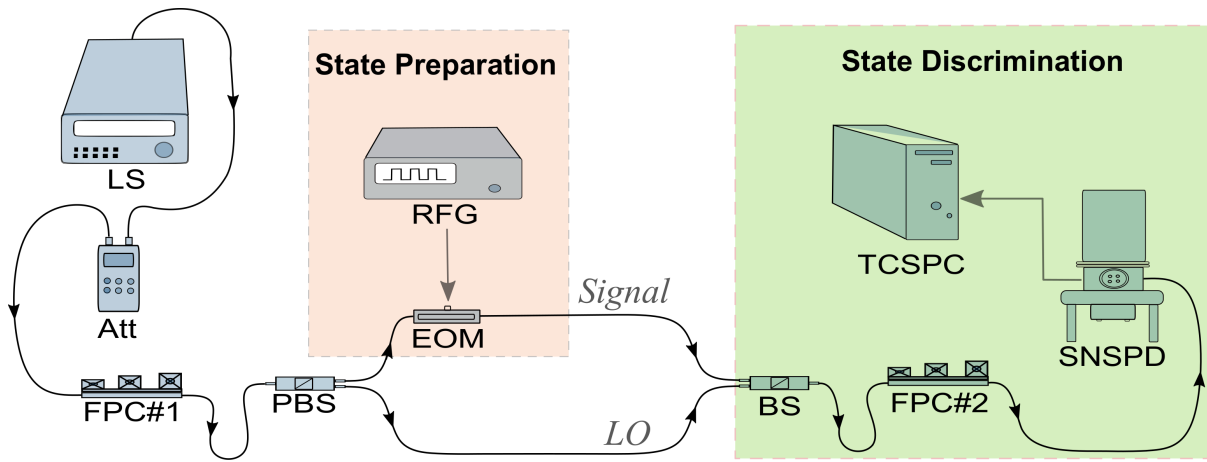}
\caption{\label{fig:setup} Schematics of the experimental setup. The source of light (LS) is attenuated via a variable attenuator (ATT) and then split into two unequal parts on a polarization beam splitter (PBS). The splitting ratio is controlled by a fiber polarization controller (FPC\# 1). The weaker part is binary phase modulated on an electro-optical modulator (EOM) driven by a radio frequency generator (RFG). The stronger part serves as a local oscillator that interferes with the signal on a beam splitter (BS). The result of interference is measured by a superconducting nanowire single-photon detector (SNSPD), preceded by a fiber polarization controller (FPC\# 2) to match the highest quantum efficiency of the detector. Time-correlated single-photon counting electronics (TCSPC) is used to synchronize the detector with the modulator.}
\end{figure*}

To realize the optical displacement, it is necessary to perform mode matching of the signal and LO. This means that all parameters describing the mode of the signal and the mode of the local oscillator must coincide. In experiment it is difficult to achieve mode matching of the spatial, temporal, spectral and polarization distributions for the light beams coming from different light sources. Thus technically both the LO and signal source are prepared by splitting one light source into two unequal parts: the larger part serves as the LO, and the smaller part is phase modulated and serves as the signal \citep{Becerra:13, Guha:11, Tsujino:11}. After mixing them on a properly balanced beam splitter, the mode-matching condition can be achieved. In fact, this approach to the implementation of optical displacement looks like an unbalanced interferometer (for example, a Mach-Zehnder interferometer), in one of whose arms we prepare the signal, and in the second arm --- the LO (see Fig.~\ref{fig:setup}). To demonstrate experimentally a quantum receiver with this approach, it is necessary to ensure stabilization of the interferometer over all freedom degrees (polarization, phase, space, timing, etc.).

In reality, it is not possible to achieve perfect mode matching and perfect stabilization of the interferometer. First, due to the non-unit interference visibility, the state $\ket{-\alpha}$ is not completely nulled. Second, non-zero dark counts contribute a certain level of background noise indistinguishable from non-perfect mode matching. Due to these factors, an additional error $e_{01}$ appears in the error diagram (see Fig.~\ref{fig:Error}).

Using formula in the work \cite{Sych:16}, the error rate accounting the interference extinction $c=I_{max}/I_{min}$ (here $I_{max}$ and $I_{min}$ are maximum and minimum intensities of the interference fringes) and dark counts $dc$ (mean number of dark counts per signal bin) of the detector reads as
\begin{equation}\label{eq:Ken2} 
e^\prime_K=0.5\* e^{-4m}+0.5 \* (1-e^{-dc-4m/c}),
\end{equation}
which is shown by red dash-dotted line in Fig.~\ref{fig:Graphics}.

To demonstrate superiority of a quantum receiver over the SQL, the system detection efficiency $\eta$, including optical loss and detector quantum efficiency, should be such that $e_{Hel}(\eta n)< e_{SQL}(n)$, which implies $\eta \gtrsim 60\%$. Theoretical error rate of the Kennedy receiver is well above the Helstrom bound, thus the sub-SQL error rate can be achieved when the minimum interference extinction $c \gtrsim 20~dB$.

\section{\label{sec:level3} Experimental setup}
Our experimental setup is shown in Fig.~\ref{fig:setup}. We use the scheme based on the Mach-Zehnder interferometer, since it allows to achieve high interference extinction and also has great potential for improving the receiver in the future. A highly coherent DFB laser operating in the CW (Continuous Wave) mode at a wavelength 1550~nm with linewidth 2 MHz is used as a light source. The power of the source is attenuated to the single-photon level by a tunable attenuator (Att). Light from DFB-laser is split by a polarization beam splitter (PBS). The splitting ratio is controlled by a mechanical polarization controller (FPC\# 1) to achieve the highest value of interference extinction. The second polarization controller (FPC\# 2) was used to adjust the polarization of the incident signal to the detector, as far as the quantum efficiency of SNSPD is polarization-dependent. 

To achieve maximum performance of the receiver, it is necessary to reduce some parasitic effects caused by optical fibers (such as polarization distortion, temperature fluctuations, vibrations, etc) that influence the stability and contrast of the interferometer. To stabilize the optical scheme, the methods described in \cite{Elezov:18, Elezov:19} were used. For polarization stability we use polarization-maintaining (PM) optical fiber components. Phase stability is achieved by increasing the thermal inertia of the optical circuit. Optical fibers were thermally stabilized by a massive metal plate placed in a sealed box.

\begin{figure}
\includegraphics[width=83mm]{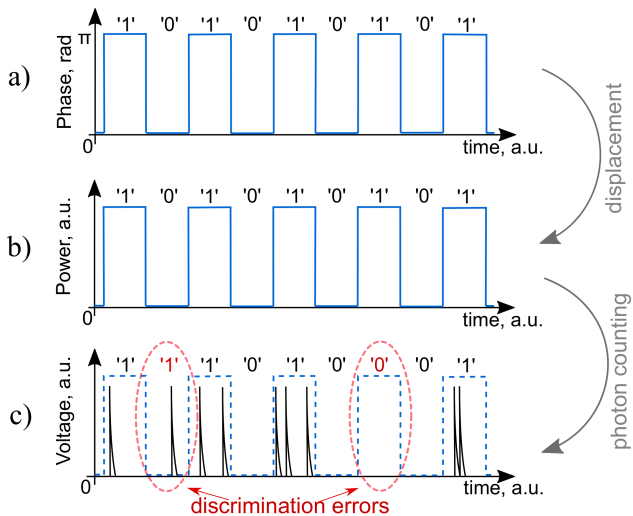}
\caption{\label{fig:Processing} Schematics of data processing. a) Initial binary data is encoded into the binary phase-shifted coherent signal. b) After optical displacement, the binary phase-shift signal is transformed into the binary amplitude-modulated signal. c) Finally, amplitude-modulated signal is registered by a single-photon detector. Photocounts are decoded into the logical data and compared with the original data. After comparison, the discrimination errors are calculated.}
\end{figure}

To estimate the discrimination error rate, it is necessary to compare the known signal at the receiver input with the decoded signal at the receiver output. Fig.~\ref{fig:Processing} illustrates the main steps we used to determine the error rate. An optical binary phase-modulated signal was generated using an electro-optical modulator (EOM). Signal parameters were controlled by a radio frequency generator (RFG). The electrical signal has a meander shape with the repetition rate 100~KHz (see Fig.~\ref{fig:Processing}a). We use this frequency since, from one hand, it allows to collect enough statistics of photocounts during the system stability time, and, from the other hand, it excludes the influence of the detector dead time (10~ns) on the statistics of photocounts. The voltage amplitude of EOM corresponds to the phase shift $\pi$. Then the signal is mixed with the LO (which has constant phase and power, matched to the signal power) that converts phase modulation into amplitude modulation (see Fig.~\ref{fig:Processing}b). The displaced signal is detected by a single-photon detector which has 65\% quantum efficiency, dark count rate about 300 counts per second, and dead time about 10~ns (SNSPD by SCONTEL)\cite{scontel}. Detection events are recorded using time-correlated single-photon counting electronics (TCSPC) with temporal resolution of 25~ps (see Fig.~\ref{fig:Processing}c). The time intervals corresponding to destructive interference encode logical ``0'', and constructive interference encode logical ``1''. The optimal selection of components for the passively stabilized optical scheme allows to achieve the interference extinction of more than 30~dB ($I_{max}/I_{min}\simeq 1250$) for the time interval of several seconds. The data acquisition time is around 1 second.

By synchronizing the RFG signal generator, phase modulator controller and TCSPC detection electronics, we can match the transmitted (Fig.~\ref{fig:Processing}a) and the received logical data (Fig.~\ref{fig:Processing}c) and determine the number of erroneously received bits. A bit is received erroneously if the half-period time window corresponding to a logical ``0'' contains some photocounts. This error $e_{01}$ corresponds to the non-zero components of the green histogram (see Fig.~\ref{fig:Error}). Also, a bit is received erroneously if the half-period time window corresponding to logical ``1'' contains no photocounts. This error $e_{10}$ corresponds to the zero component of the blue histogram (see Fig.~\ref{fig:Error}).

\section{\label{sec:level4} Experimental results}

\begin{figure}[h]
\includegraphics[width=85mm]{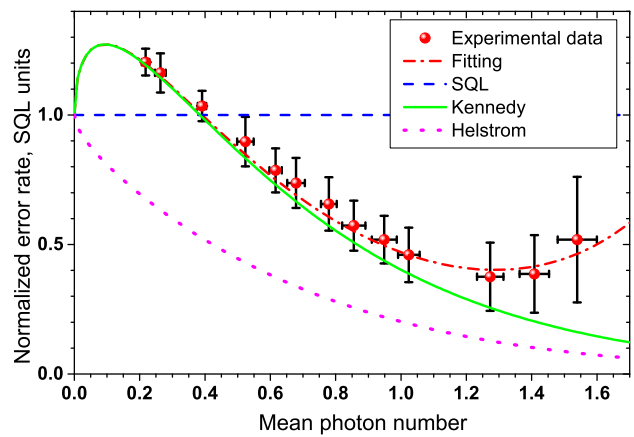}
\centering\caption{\label{fig:Graphics} Discrimination error rate as a function of mean photon number in a signal time bin. The error rate is normalized to the standard quantum limit (SQL), shown by the blue dashed line. Red dots show experimental error rate of our receiver, including the statistical deviations indicated by black bars. Red dash-dotted line shows theoretical fit for the measured error rate. For comparison, we also show the theoretical values for the ideal Kennedy receiver (green solid line) and the Helstrom bound (magenta dotted line).}
\end{figure}

Using experimental data, we calculate the error rate for different levels of signal intensities, measured in mean number of photons per signal time bin (or, simply, number of photons). The results are presented as red points in Fig.~\ref{fig:Graphics}. All data points in Fig.~\ref{fig:Graphics} are normalized to the SQL (Eq.~\ref{eq:SQL}). The mean photon number is derived from the mean number of photocounts, i.e. we do not take into account optical losses (0.3 dB on the beam splitter) and quantum efficiency of SNSPD (65\%). The aim of this work is a proof-of-principle demonstration of the all-fiber Kennedy receiver. Technically, we could fuse two fibers and eliminate the optical loss of the beam splitter connectors, as well as replace SNSPD by a more efficient one, available at SCONTEL.

For comparison, we show the Helstrom bound (Eq.~\ref{eq:Hel}) by the magenta dotted line, the error rate of the ideal Kennedy receiver (Eq.~\ref{eq:Ken}) by the green solid line, and the error rate of the non-perfect Kennedy receiver (Eq.~\ref{eq:Ken2}) by the red dash-dotted line. The last curve fits well the experimental points for the dark count rate of $\sim 300$~Hz, which corresponds to dark count probability per signal bin $\sim 1.5 * 10^{-3}$.

According to the obtained results, Kennedy receiver has high sensitivity in the range between 0.5 and 1.6 photons. The experimentally observed error rate in this range is well below the SQL. The lowest error rate, normalized to SQL, was obtained for the signal intensity 1.3 photons and is equal to 0.4 SQL.

For the signal intensities less than 0.4 photons, performance of the Kennedy receiver is worse than SQL due to the $e_{10}$ error (the first component of the blue histogram in Fig.~\ref{fig:Error}). This is the fundamental limitation of the Kennedy receiver. For signal intensities higher than $1.6$ photons, performance of the Kennedy receiver is also worse than SQL, but this is a technical issue. In this case, the dominant contribution is $e_{01}$ error (the non-zero components of the green histogram in Fig.~\ref{fig:Error}). It originates from the non-unit interference visibility and dark counts, which set the background noise. The higher intensity of the signal, the larger this contribution, hence higher discrimination error rate. To achieve high sensitivity of the receiver in this domain, we need to observe high interference extinction, which is a difficult task for practical implementation. Further progress in this direction may be associated with an improvement of the interference extinction, overall stability of the optical circuit and development of novel receiver designs.

\section{\label{sec:Con} Conclusions and outlook}

We experimentally demonstrated all-fiber quantum receiver based on Kennedy's design. We used a two-arm polarization-maintaining optical fiber Mach-Zehnder interferometer to create a binary phase-shift keyed signal in one arm and local oscillator in the other arm. Also, we used a very promising superconducting nanowire single-photon detector with excellent characteristics to measure the displaced signal. We achieved interference extinction above 30dB, which allows us to observe discrimination error rate below the SQL for the signal intensity level between 0.5 and 1.6 photons per time bin. The minimum value of error rate that we experimentally obtain is 60$\%$ below the SQL. To the best of our knowledge, this is the first completely all-fiber realization of a quantum receiver with sub-SQL performance.

The main motivation for this work is to demonstrate a quantum receiver solely based on standard fiber-optic components. The use of an optical fiber circuit makes it easier to employ the receiver in practical systems, such as quantum key distribution with discrete modulation of coherent states \cite{Sych:10}. Our work is also aimed at creating a receiver with the potential for its further modernization and development of more advanced schemes \cite{Sych:16}. We hope that the practical implementation of quantum receivers with high sensitivity can give impetus to their wider use in related research fields.

\begin{acknowledgments}
We thank Scontel for providing a single-photon detector. The research has been carried out with the support of the Russian Science Foundation (project No. 17-
72-30036)
\end{acknowledgments}

\bibliography{Library_receiver}

\end{document}